\documentclass[preprint2]{aastex61}
\usepackage[utf8]{inputenc}

\usepackage{natbib}
\usepackage{graphicx}

\shorttitle{Misaligned USPs in STIPs}
\shortauthors{Becker et al.}

\begin{document}

\title{The Origin of Systems of Tightly Packed Inner Planets with Misaligned, Ultra-Short-Period Companions}

\correspondingauthor{Juliette Becker}
\email{jbecker@caltech.edu}

\author[0000-0002-7733-4522]{J. Becker}
\altaffiliation{51 Pegasi b Fellow}
\affiliation{Division of Geological and Planetary Sciences, Caltech, Pasadena, CA}
\author[0000-0002-7094-7908]{K. Batygin}
\affiliation{Division of Geological and Planetary Sciences, Caltech, Pasadena, CA}
\author[0000-0003-3750-0183]{D.~Fabrycky}
\affiliation{Department of Astronomy and Astrophysics, University of Chicago, Chicago, IL}
\author[0000-0002-8167-1767]{F.~C.~Adams}
\affiliation{Department of Physics, University of Michigan, Ann Arbor, MI}
\author[0000-0001-8308-0808]{G. Li}
\affiliation{School of Physics, Georgia Institute of Technology, Atlanta, GA}
\author[0000-0001-7246-5438]{A.~Vanderburg}
\altaffiliation{NASA Sagan Fellow}
\affiliation{Department of Astronomy, University of Texas, Austin, TX}
\affiliation{Department of Astronomy, University of Wisconsin-Madison, Madison, WI}
\author[0000-0001-8812-0565]{J.~E.~Rodriguez}
\affiliation{Harvard Astronomy Department, Cambridge, MA}

\begin{abstract}
Ultra-short period planets provide a window into the inner edge of the parameter space occupied by planetary orbits. In one particularly intriguing class of multi-planet systems, the ultra-short period planet is flanked by short-period companions, and the outer planets occupy a discernibly distinct dynamical state. In the observational database, this phenomenon is represented by a small number of stars hosting systems of tightly packed co-planar planets as well as an ultra-short period planet, whose orbit of is misaligned relative to the mutual plane of the former. In this work, we explore two different mechanisms that can produce an ultra-short period planet that is misaligned with the rest of its compact planetary system: natural decoupling between the inner and outer system via the stellar quadrupole moment, and decoupling forced by an external companion with finely-tuned orbital parameters. These two processes operate with different timescales, and can thus occur simultaneously. In this work, we use the K2-266 system as an illustrative example to elucidate the dynamics of these two processes, and highlight the types of constraints that may arise regarding the dynamical histories of systems hosting ultra-short period planets.

\end{abstract}

\section{Introduction}
Among the thousands of planet discoveries made by the Kepler and K2 missions, a significant fraction of them exhibit a surprisingly ubiquitous system architecture: systems of tightly packed inner planets (known as STIPs). This classification denotes systems that contain multiple planets orbiting with small semimajor axes (generally with $a<0.5$ AU). The high occurrence rate for these compact systems, estimated to be $\sim$20-30\% \citep{Muirhead2015,Zhu2018}, suggests that this geometry is one of the dominant outcomes of the planet formation process. 

Additional observations of these compact systems show that they generally exhibit a surprising degree of regularity. The systems are discovered with multiple transiting planets and seem to be co-planar; more specifically, their inclination angles appear to be drawn from a relatively narrow distribution \citep{Fabrycky2014}. The planetary orbits in these systems display uniform orbital spacing \citep{Rowe2014} with low eccentricities \citep{Xie2016, Mills2019}. In addition, the planets in these systems tend to have {far more intra-system regularity than would result from random samples of the entire exoplanet mass distribution, with both similar radii \citep{Weiss2018} and masses \citep{Millholland2017}}. 

Ultra-short-period (USP) planets - often defined as having periods less than a day - represent a less common type of object. These extreme objects occur around $\sim$0.5\% of G-dwarf stars \citep{SanchisOjeda2014} and provide another remarkable challenge to the standard model of planet formation.

The intersection of these two sets (STIPs and USPs) has recently showed an intriguing geometry: STIPs are not always coplanar, particularly when one member of the STIP is a USP. 
{High-multiplicity systems of mostly coplanar systems are generally thought to form via disk migration \citep{Hansen2012, Rein2012}, particularly when some of those coplanar planets are in or near resonance \citep{Papaloizou2003, Baruteau2013, Batygin2015mnras, Deck2015}. }
Two recent discoveries{show systems where high multiplicity systems only partially show the coplanarity expected from disk migration}: K2-266, a five-planet STIP {with two near-resonant planets} which has an inner USP misaligned {by an orbital inclination difference of} 14 -- 17 degrees \citep{Rodriguez2018}; and TOI-125, a four-planet STIP with an inner USP misaligned by 16 -- 20 degrees \citep[however, we note that the TOI-125 USP is a planet candidate with a low SNR and has not yet been confirmed;][]{Quinn2019}. Both of these systems simultaneously conform to the STIP archetype (as all planets except the inner one are very tightly packed and nearly coplanar to each other) as well as subvert it (the USP in each system is significantly misaligned compared to the outer plane of planets).

The results of \citet{Dai2018} show that planets with ultra-short orbital periods populate a larger fraction of the `transiting' range of inclinations than do planets with slightly longer periods. 
In inclination space, the outer system is more planar, while inner systems in comparison tend to be more extended and fill the full allowable transiting parameter space. 
K2-266 fits this paradigm: K2-266 b, with its grazing transit, resides at the absolute maximum mutual inclination that is still observable, while the outer system is very compact in vertical space. 

We remark that the question of the underlying distribution of planetary inclination in transiting systems is really a question of the underlying populations of planets. The Kepler dichotomy \citep{Johansen2012} --- an apparent mismatch (in several observational dimensions; \citealt{Morton2014, Ballard2016}) between multi- and single-planet systems --- might probe either an observational effect or an underlying property of planet formation \citep{Moriarty2016}. 
\citet{Spalding2016} suggest that misalignments of stellar spin axes can explain the Kepler dichotomy by generating systems of multiple planets with mutual misalignments, resulting in some multi-planet systems being mistakenly interpreted as single-planet systems while in other cases the spin-orbit misalignment can drive a dynamical instability. 
Simultaneously, the existence of systems like Kepler-56 \citep{Huber2013} with high stellar obliquity, multiple well-aligned {(defined as low inclination dispersion)} transiting planets, and a long-period massive planet suggests that the secular dynamics of an exterior companion could cause the observed misalignment between the inner planets and the stellar spin axis in this particular system \citep{Boue2014a, Boue2014b, Gratia2017}. Similar dynamics would be at play in any system with a STIP and an exterior, inclined companion \citep{VanLaerhoven2012, Hansen2017, Mustill2017, Becker2017, Jontof2017, Denham2019, Masuda2020}. 
It is worth noting, however, that giant planets are likely less common than STIPS, implying that such a process is probably effective in a minority of systems at a population level. 

The USPs in the TOI-125 and K2-266 systems reside at nearly the largest impact parameters at which they can still be observed {while the other planets have inclinations close to 90 degrees as derived from photometry and TTVs}, thus posing the question: how many existing STIPs have unseen, misaligned USP companions? The answer to this question would shed light on the outcomes of planet formation that lead to the formation of USPs and the physical processes that underlie the Kepler dichotomy.

In this work, we consider two possible mechanisms that could explain how STIPs can be observed to host interior, misaligned {(by an amount several times greater than would be expected from the typical \citealt{Fabrycky2014} Rayleigh distribution used to describe STIP inclinations,)} USPs and still maintain mutual inclinations {within a few degrees} among the outer (non-USP) planets. In Section \ref{sec:archetecture}, we discuss the architecture of this class of systems and describe why an additional effect is needed to explain the present-day geometry. We then present two mechanisms that could resolve this issue: The first mechanism, discussed in Section \ref{sec:j2}, requires an oblate central star with a slight amount of stellar obliquity with respect to the original plane of planets. The second mechanism, discussed in Section \ref{sec:companion}, requires the existence of an additional (as yet unseen) planet in the system. Both of these mechanisms can reproduce the observed effect. In Section \ref{sec:conclude}, we compare the regimes in which the two mechanisms can operate and  conclude with a summary of our results.

\section{Current-Day Dynamics of a STIP with a Misaligned USP}
\label{sec:archetecture}

\begin{figure}
 \includegraphics[width=3.2in]{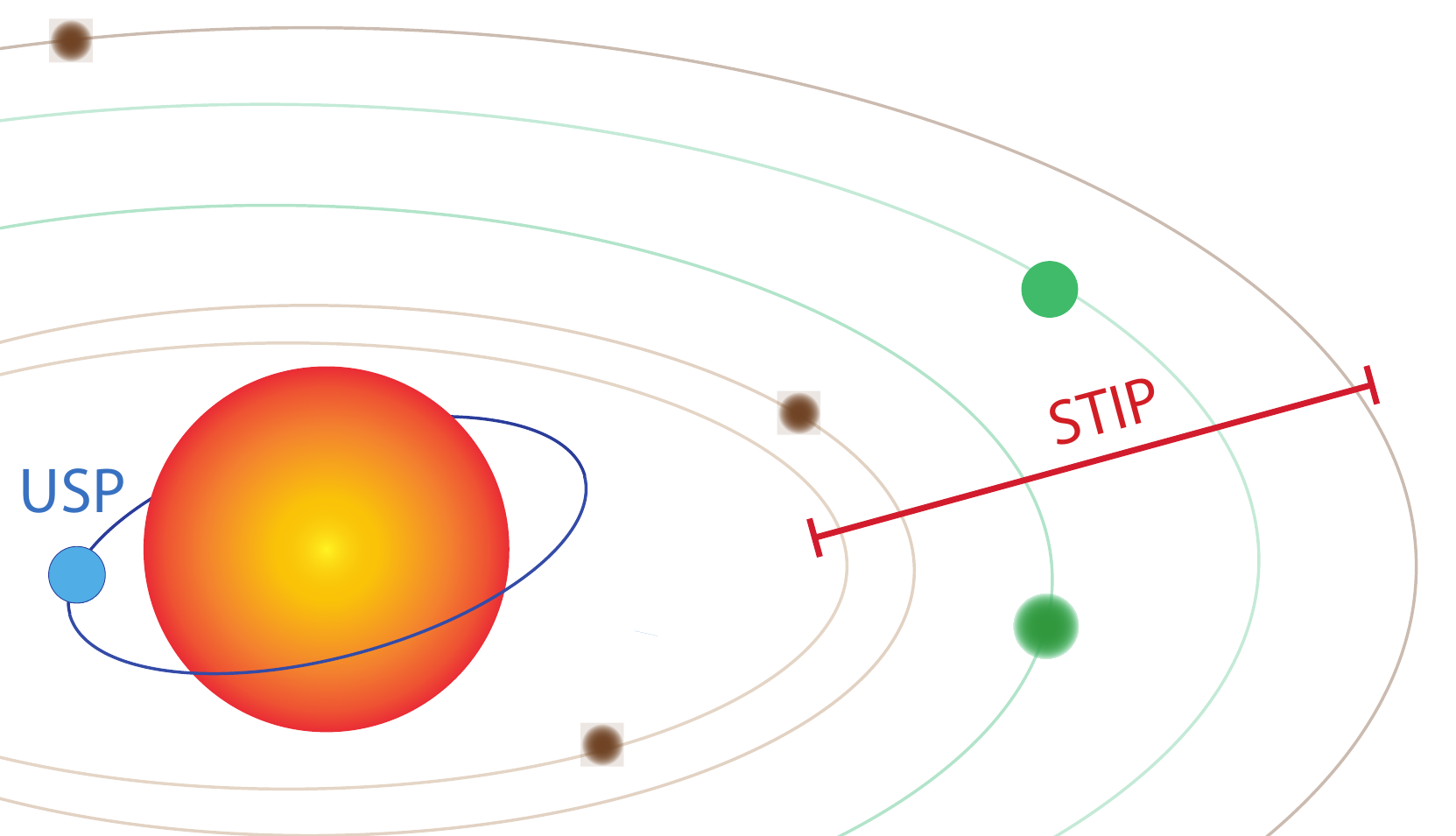}
 \caption{A schematic of the geometry of planetary system, discovered to exist in TOI-125 and K2-266, and considered in this work.  Systems of this type are composed of two components: an inner ultra-short period planet, and an outer plane of tightly-packed, multiple planets, which have low mutual inclinations with each other. Between the USP and the outer plane, there is a significant ($\sim 15$ degree or greater) misalignment. The specific geometry portrayed in this schematic is the K2-266 system \citep{Rodriguez2018}.}
    \label{fig:stip}
\end{figure}

Results from the Kepler Mission indicate that STIPs are a natural end state of the planet formation process, while USPs are considerably more rare {(occurring in 20-30\% vs. 0.5\% of systems respectively; \citealt{SanchisOjeda2014,Muirhead2015, Zhu2018})}. In K2-266, the low mutual inclinations of the outer STIP is maintained, even as the USP is misaligned significantly compared to the outer plane of planets. This system geometry under consideration is illustrated in Figure \ref{fig:stip}. 
{In the K2-266 system, the innermost planet has a measured inclination of 75.3 $\pm$ 0.6 degrees, while the outer planets have inclinations ranging between 
87.8 and 89.5 degrees (with errors ranging from 0.1 to 0.8 degrees). The difference between the USP and the rest of the planets is not accounted for by the observational errors, and instead shows a physical misalignment in inclination.}


Combining a STIP with a misaligned USP requires the angular momentum transfer among the STIP to be rapid compared with the STIP-USP interactions {to ensure that the STIP does not attain a configuration with enough mutual inclination so that the planets would not be seen to be transiting {\citep[for a secular treatment of angular momentum exchange in exoplanet systems, see][]{MD99}}}. When observing the evolution of the mutual inclinations, two main features must be preserved:
\begin{itemize}
    \item The USP must be misaligned {with respect to the mean orbital plane of} the outer STIP.
    \item The mutual inclination of the {planets in the STIP} must remain low {such that the planet can be seen in transit}.
\end{itemize}
To form a system like K2-266, the dynamical evolution of the system must produce the first condition while still maintaining the second. In this work, we focus on K2-266 and its specific parameters; however, our analysis can be applied to any STIP with an inner, misaligned USP. 

As shown by \citet{Rodriguez2018}, once the misalignment between the USP and STIP in K2-266 has been produced, it is naturally maintained through secular interactions.  {\citet{Rodriguez2018} did not consider tidal interactions, and the current-day planet parameters suggest that the tidal timescale for the inner planet is much longer than the planet-planet interaction timescales, so we also do not consider tides in this work. }
In this work, we consider what system properties will create the initial misalignment that allows the dynamics we see today in the K2-266 system (and by extension, other similar systems like TOI-125). 
The dynamics of this particular system are affected by resonance (the 14- and 19-day period planets are close to or in the 3:4 resonance) as well as long-term secular evolution. As such, in our analysis we use numerical N-body integrations to model the dynamics in order to include all of these effects. 

\subsection{Global numerical integration parameters}
\label{sec:sim_params}
We use the same basic parameters for all simulations, altering only the stellar obliquity, stellar $J_{2}$ moment, and (at times) introducing an additional (so far undetected) outer planetary companion with varying parameters. 
First, we use a single set of initial conditions for the known planets (see Table \ref{tab:system_parameters}). We include only the validated planets (K2-266 b, c, d, e), ignoring the candidate planets (K2-266.02 and .06). 
{The candidates .02 and .06 were not validated in \citet{Rodriguez2018} because their signal to noise ratio was too low \citep[see validation procedure in][]{Morton2016}. We used a small set of test simulations to assess whether the presence/absence of these two non-validated planets from the simulations significantly impacts our results, and found that .06 does not alter the results, and the effect of planet 0.02 can result in dynamical instability at high J2 values (for some values of its orbital elements). Since we do not study dynamical stability in this work, and 0.02 is not validated, we leave the consideration of (if real) what additional information we could learn from it to future work and proceed only with the validated planets.}

For all star and planet physical and orbital parameters, we have taken the median value of the EXOFASTv2 \citep{Eastman2017, Eastman2019} posterior generated by the fit in \citet{Rodriguez2018}. 
To further simplify the system, we also set all {initial} planetary eccentricities to zero. We also set the starting planetary inclinations to be equal to each other. {By default, these are set to be 0 degrees in the simulation frame, which is equivalent to 90 degrees in the transit-fit frame.} These inclinations are not the values seen in the current system (where the USP is misaligned relative to the other four). For the simulations, we set all planets to start in the same orbital plane because we are attempting to determine how the USP-STIP misalignment arose, assuming that the planets started roughly coplanar in the protoplanetary disk and reached their final orbital positions in roughly that configuration. 
{Since the purpose of this set of simulations is to examine the reliance of the dynamics on a single parameter (the stellar $J_2$), to minimize the number of points we needed to run at each $J_2$, we did not start with any inclination dispersion among the planets (which would add an envelope of noise to our results).}
If the secular evolution of the planets, originating from the current state, does not return the planets to {roughly} $i\sim0$, then secular dynamics alone cannot produce misalignment without an additional perturbation, as the current system and expected initial condition are not on the same dynamical trajectory. 

\begin{table*}
  \begin{center}
    \caption{The values of parameters used for simulations in this work. Values are median draws from \citet{Rodriguez2018}. Note that to reduce the number of degrees of freedom for our simulations, we do not draw values from the entire posterior, but instead take only the median values. Orbital eccentricities are set to zero ($e=0$). {$\omega$ denotes the planetary argument of pericenter. } }
    \label{tab:system_parameters}
    \begin{tabular}{p{2cm}p{2cm}p{3cm}p{1cm}p{1cm}p{1.5cm}p{1.5cm}p{2cm}} 
\multicolumn{2}{l}{{Planetary Parameters}}			&		&		&		&		&		\\
{Planet}	&	{Orbital Period (days)}	&	{Time of conjunction (BJD)}	&	{Ecc. $e$} & {$\omega$ (deg)}	&	{Radius ($R_{\oplus}$)}	&	{Mass ($M_{\oplus}$)}	&	{Included in simulations?}	\\
K2-266 b	&	0.6585	&	2457945.719	&	0	&0	&	3.3	&	11.30	&	Yes	\\
K2-266.02	&	6.0993	&	2457913.42	&	0	&88	&	0.646	&	0.21	&	No	\\
K2-266 c	&	7.8204	&	2457907.543	&	0 &	87	&0.705	&	0.29	&	Yes	\\
K2-266 d	&	14.6971	&	2457944.842	&	0	&87	&	2.93	&	9.40	&	Yes	\\
K2-266 e	&	19.4819	&	2457938.54	&	0	&89	&	2.73	&	8.30	&	Yes	\\
K2-266.06	&	56.7024	&	2457913.419	&	0	&83	&	0.9	&	0.70	&	No	\\
			&		&		&		&		&		\\
\multicolumn{2}{l}{{Stellar Parameters}}			&		&		&		&		&		\\
\multicolumn{2}{l}{Stellar Mass		}	&	0.686 $M_{\odot}$	&		&		&		&		&		\\
\multicolumn{2}{l}{Stellar Radius	}		&	0.703 $R_{\odot}$	&		&		&		&		&		\\
\multicolumn{2}{l}{$J_{2}$	}		&	Variable	&		&		&		&		&		
    \end{tabular}
  \end{center}
\end{table*}

For all simulations, we use the \texttt{Mercury6} \citep{Chambers1999} N-body integrator in hybrid symplectic and Bulirsch--Stoer (B-S) mode with an initial time-step of 7.2 minutes, which is roughly 0.75\% of the shortest dynamical time in the system. Integration lengths are set to $10^{5}$ years and energy was conserved to better than one part in $10^{9}$ (for energy changes due to the integrator). We include {first-order post-Newtonian} general relativistic precession due to the central body only. {Note that while the system's short dynamical time and limited computational resources require a limited integration time, longer-term dynamical effects might modify the results the presented here but are the subject of future work.}

In the remainder of this manuscript, the simulation parameters enumerated above are used except when specified otherwise.

\section{Effect of Stellar Oblateness on USP-STIP interactions}
\label{sec:j2}
In addition to planet-planet interactions, secular dynamics of planetary systems can be affected by stellar oblateness. The higher order moments of the star's gravitational potential are often described by $J_{N}$, dimensionless constants that scale as follows: $J_{N} \propto (\Omega^{2} R_{*}^{3} G^{-1} m_{p}^{-1})^{N/2}$, where $N$ is the term of the potential being considered, {$R_{*}$ the stellar radius, $G$ the gravitational constant, $m_{p}$ the planetary mass, and $\Omega$ the stellar angular velocity}. The quadrupole moment, $J_{2} \sim m_{p}^{-1}$, is the dominant term in this series. As reference point, the {first moment for the Sun is}  $J_{2} \approx 2\times10^{-7}$ \citep{Duvall1984,Mecheri2004}. These harmonics describe the effect of rotational deformation, which has a particularly large effect on {the orbits of} ultra-short-period planets.  

The idea that stellar oblateness affects the orbits of nearby planets has been suggested as a possible solution for the Kepler dichotomy \citep{Spalding2016} as well as an explanation \citep{Li2020} for why shorter period exoplanets tend to have larger spreads in orbital inclinations \citep{Dai2018}. 
\vspace{5mm}


\subsection{Systems around an oblate star with a slightly misaligned stellar spin axis}
\label{sec:misaligned_axis}
Just as the Sun has a slight {6 degree} obliquity relative to the mean plane of the Solar System \citep{Souami2012}, exoplanet-host stars can have some obliquity with respect to their planetary companions.  Although low-mass stars that host STIPs tend to have lower stellar obliquities {than their high-mass counterparts} \citep{Morton2014}, their values remain nonzero. Moreover, observed stellar obliquities often exceed the spread of mutual inclinations among the planets. A stellar obliquity up to about 20 degrees can arise naturally \citep{Batygin2013, Spalding2015}. In this section, we assume that the star hosting the USP-STIP system has some {non-zero} natal obliquity relative to the planetary orbital planes, and model how the inclination evolution depends on the stellar $J_{2}$ moment.

\begin{figure*}
 \includegraphics[width=7in]{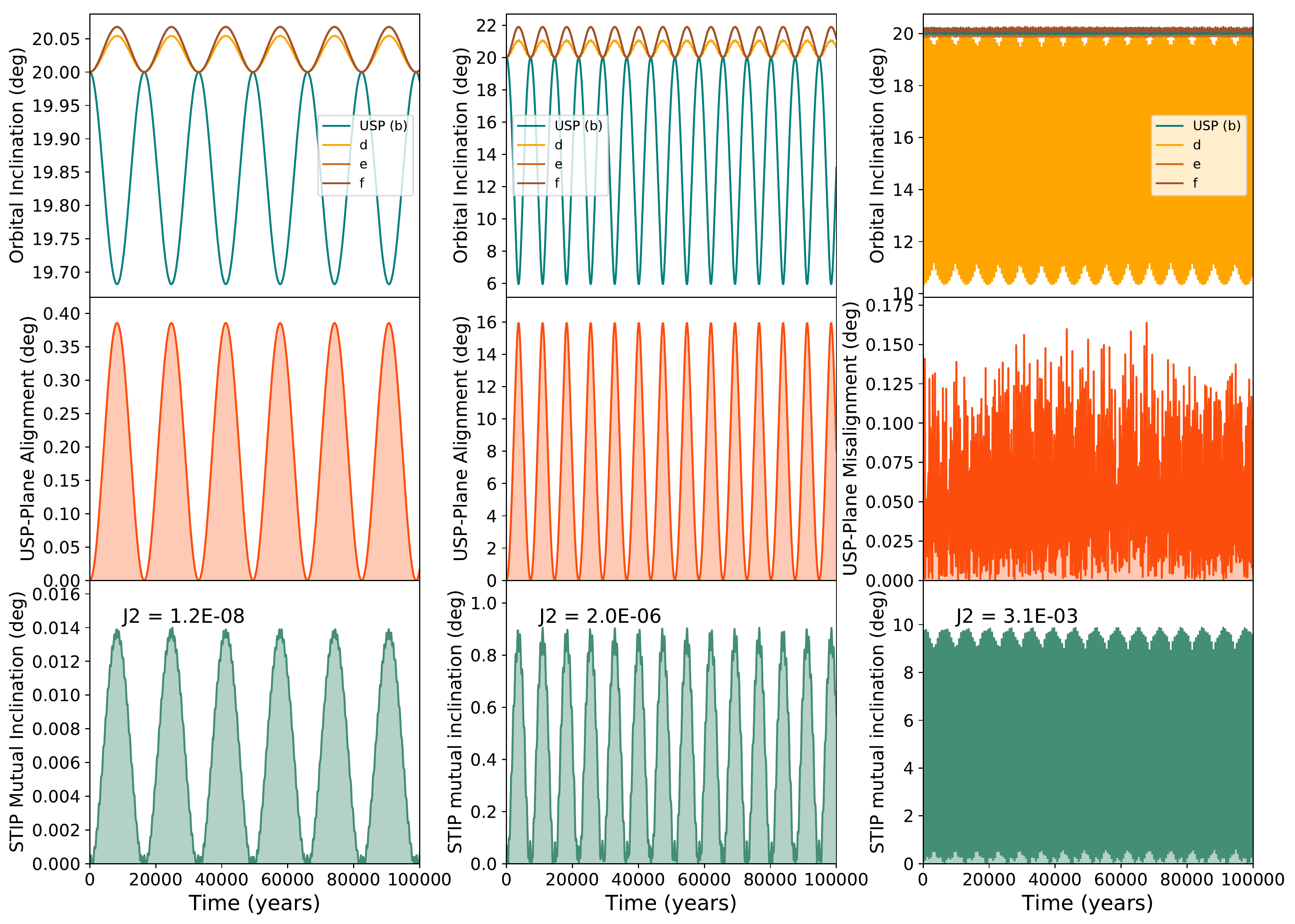}
 \caption{The N-body evolution of the four validated/confirmed planets in the K2-266 system under three different (constant) values of the stellar quadrupole moment $J_2$ (where the stellar spin axis is misaligned by 20 degrees relative to the plane containing all four planets). The three panels show (top panel) the evolution in inclination for each planet, (middle panel) the misalignment between the USP and the outer plane of planets (where the outer plane location is taken to be the median inclination), and (bottom panel) the {inclination dispersion} of the STIP {(defined as the mathmetical range of inclinations of all planets excluding the USP)}. In the left column, $J_2 = 1.2 \cdot 10^{-8}$, and all planets (the USP and the outer members of the STIP) remain well-confined with low mutual inclinations); in the middle column, $J_2 = 2.0 \cdot 10^{-6}$, and while the outer planets keep a low mutual inclination, the USP is highly misaligned relative to the outer STIP; in the left column, $J_2 = 3.1 \cdot 10^{-3}$, and all planets have higher mutual inclinations than we see in a typical STIP. {The amplitude of oscillations in inclination can be approximately analytically described by secular theory outlined in \citet{MD99}.}  }
    \label{fig:part1}
\end{figure*}

Using the simulation and system parameters described in Section \ref{sec:sim_params}, we integrate the system 150 times with values of $J_{2}$ varying between $10^{-10}$ and $2 \cdot 10^{-3}$. {For the trials considered in this section,} we set the initial misalignment between the stellar spin axis and planetary orbital angular momenta directions to be 20 degrees. To perform these simulations, we use the Open Science Grid \citep[OSG;][]{osg1} accessed through XSEDE \citep{xsede}. Three selected trials with very different values of $J_{2}$ are shown in Figure \ref{fig:part1}, which demonstrates differences depending on $J_{2}$ in (a) the planetary inclination evolutions, (b) the measured amplitude of the misalignment between the USP and the median plane of the STIP, and (c) the {mutual inclination between all members of the STIP}. 

Despite our simulations starting with a simplified, idealized version of K2-266's initial conditions, not all choices of stellar $J_{2}$ create systems {qualitatively} consistent with K2-266's current geometry. Recall, the goal of this analysis is to find a mechanism that both (a) creates the misalignment between the USP and the STIP components of the system, but also (b) keeps the mutual inclination in the system low. From the three columns in Figure \ref{fig:part1}, {it is} clear that if $J_{2}$ is too low, the misalignment between the USP and STIP is too small; if $J_{2}$ is too large, then the spread in inclination of the STIP becomes too large so that not all of the outer planets would be seen to transit at the same time. 
\begin{figure}
 \includegraphics[width=3.4in]{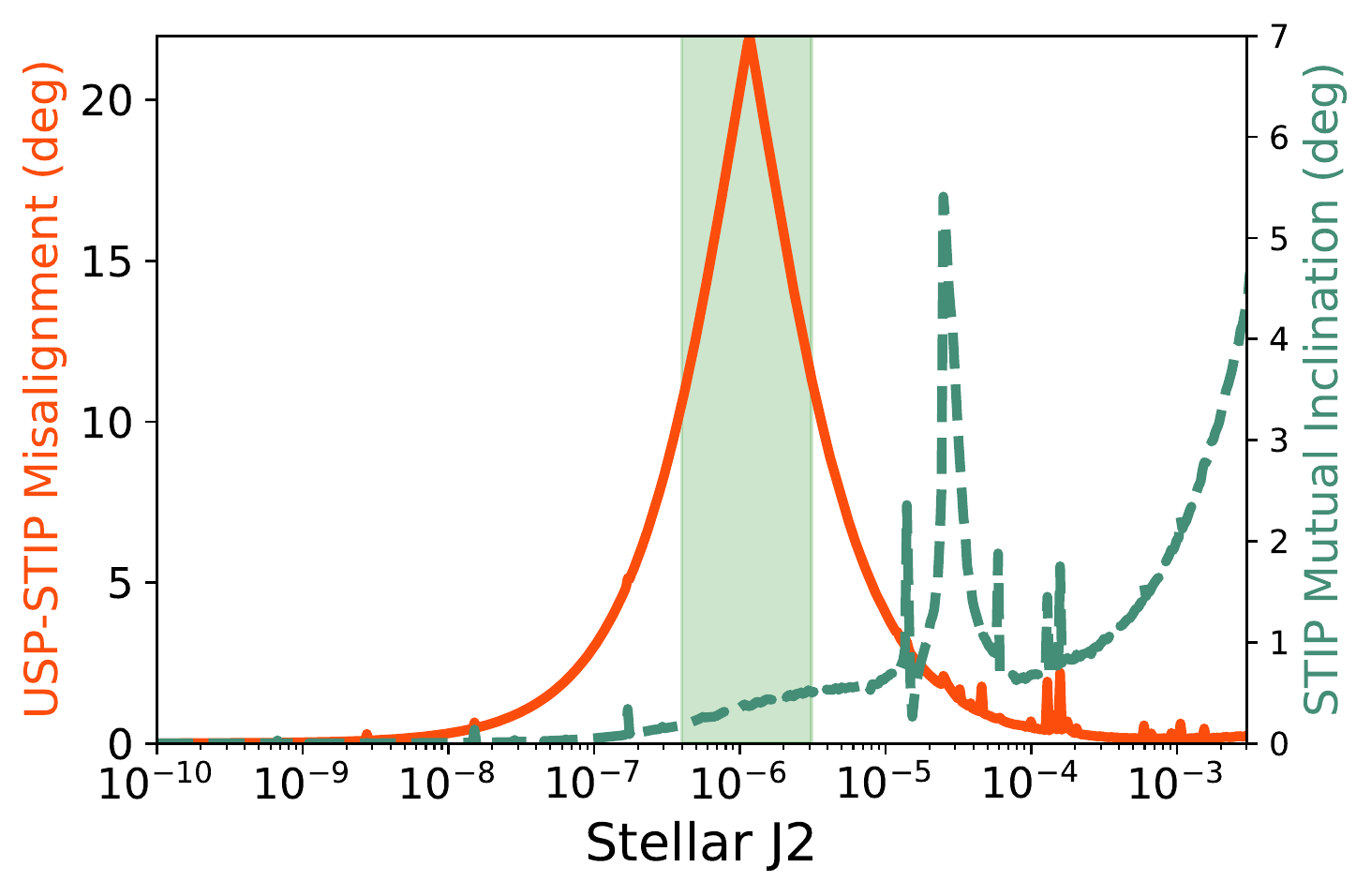}
 \caption{The maximum misalignment between the USP and the outer STIP (left scale) and the median spread in the mutual inclination of the outer STIP (right scale) as a function of stellar $J_2$. The host star was given an initial obliquity of 20 degrees with respect to the plane of planets. The shaded region denotes the approximate regime in which both criteria for reproducing the observations are satisfied, {defined as geometries where the maximum USP-STIP misalignment observed exceeds the value observed in the K2-266 system today while the STIP maintains a transiting configuration.} }
    \label{fig:metrics}
\end{figure}

The results from all 150 simulations are shown in Figure \ref{fig:metrics}, which plots the two metrics (USP misalignment and STIP confinement in inclination space) as a function of $J_{2}$. We find that for our initial stellar obliquity of 20 degrees, and for intermediate values of $J_{2}$ (more specifically, between $5\cdot10^{-7}$ and $5\cdot10^{-6}$), we satisfy both metrics and create systems with geometries like that of K2-266. 

We note that the analysis here neglects the dynamics induced by a $J_{2}$ evolving {in time.} Instead, what we show here is for a range of possible values of $J_{2}$, what happens to a system that arrives at their present-day orbital positions with no mutual inclination while the star has that particular value of $J_{2}$. A star with a value of $J_{2}$ in that range could thus produce the observed misalignment on relatively short ($<1$ Myr) timescales, while allowing the outer plane of planets to remain compact.

\subsection{Varying the initial stellar obliquity}
The next natural question is how the magnitude of the initial stellar misalignment affects this result. The simulations in the previous section were initialized with an initial stellar obliquity of 20 degrees.

We now repeat the analysis of the previous section, but use different values for the initial stellar obliquity. Again, we use the base simulation parameters described in \ref{sec:sim_params}. We have run 3 suites of 150 simulations, each with a different value for the initial stellar obliquity: $\psi = (3, 10, 20)$ degrees. We vary the $J_{2}$ in each of those 150 simulations as before, and keep all other parameters the same as in the previous section.  

\begin{figure*}
 \includegraphics[width=7.in]{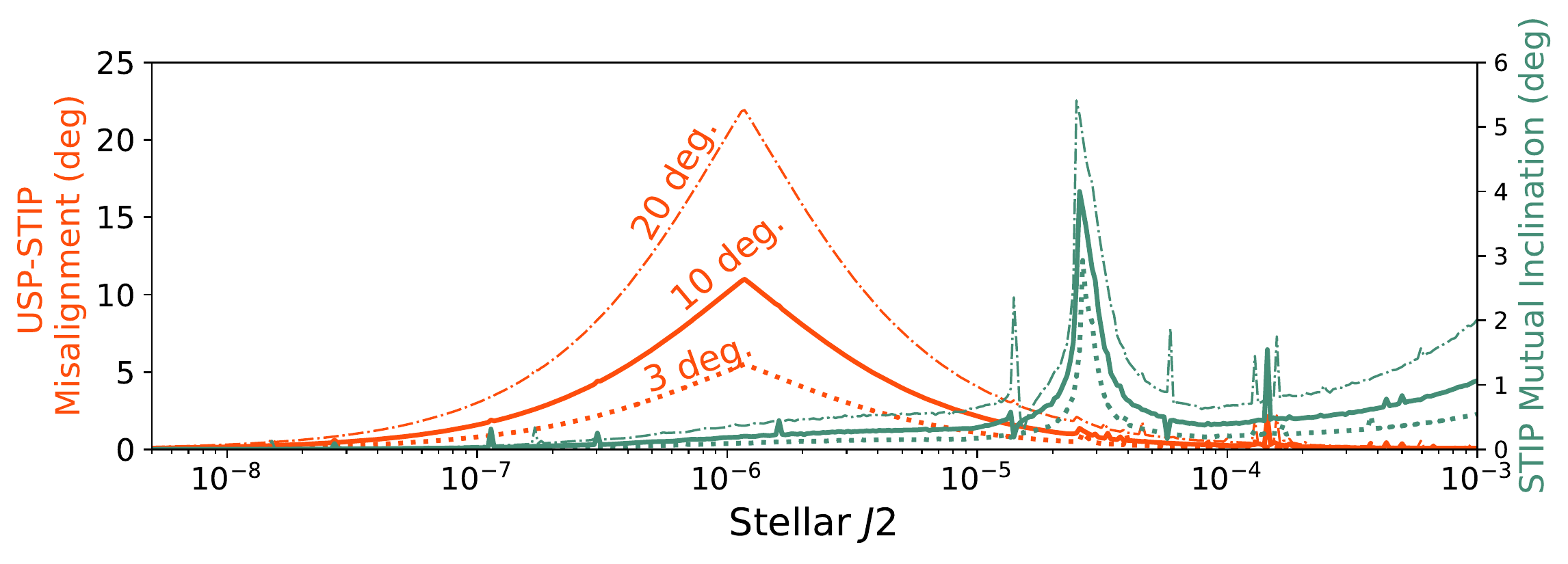}
 \caption{A version of Figure \ref{fig:metrics} including three sets of simulations, varying only in the initial stellar obliquity assigned to the star. From top (dot-dash) line to bottom (dotted) line, the initial obliquities are 20 degrees, 10 degrees, and 3 degrees. The peaks in $J_{2}$ space occur in the same locations (locations which are defined by the orbital periods of the planets rather than the angular momentum-related parameters of the star), but with different heights depending on the initial stellar obliquity. }
    \label{fig:metric2}
\end{figure*}

The results of this analysis are shown in Figure \ref{fig:metric2}, an expanded version of Figure \ref{fig:metrics}. There are three important conclusions from this analysis. First, as the stellar obliquity is decreased, the USP-STIP misalignment also decreases. Second, as the stellar obliquity is decreased, the mutual inclination of the outer STIP also decreased, i.e., a more tilted star causes more disruption in mutual inclinations. Third, the peaks (in misalignment and inclination spread) occur at particular values of $J_{2}$ for all initial obliquities. The exact locations of these peaks are due to the orbital properties of the planets {(primarily the ratios of the semi-major axes)} rather than the obliquity of the star. If we were to remake this diagram for a different system, the peaks would be in different locations depending on the secular resonance locations in that other system. As such, the specific quantitative results we find for K2-266 in this work are not universal, but must be recomputed for the particular architecture of other systems.   


\subsection{The Evolving Host Star}
As shown in Figure \ref{fig:metrics}, large values of $J_{2}$ cause large mutual inclinations to be excited between the planets composing the STIP. We know already from other observational constraints \citep{Li2020} that stars such as K2-266 do not currently have such large values of $J_{2}$. However, as the stars descended onto the main sequence, they are likely to have passed through all values of $J_{2}$ relevant to this work. We can use simple scaling relations combined with stellar evolution models to determine the ages at which K2-266 would have passed through each of these milestones. The goal of this analysis is to determine which values of $J_{2}$ we can reasonably expect these planets to have experienced if they formed while the protoplanetary disk was still present. 

To estimate the time evolution of $J_{2}$, first we define $J_{2}$ in terms of the angular velocity $\Omega$, break-up angular velocity $\Omega_b$, and stellar love number $k_2$, i.e., 
\begin{equation}
J_{2} = \frac{1}{3}(\frac{\Omega}{\Omega_b})^{2} k_{2} \,. 
\label{eq:scaling}
\end{equation}

As a star descends onto the main sequence, two things can happen that would affect the dynamics we describe above: first, the magnitude and direction of the stellar spin axis' obliquity with respect to the protoplanetary disk can change \citep{Batygin2013}; second, the value of $J_{2}$ will decrease over time. This section considers the second process in isolation, and leaves the first for future work. {As a result, this work can be applied to systems where the planets form further out in the disk and migrate to their observed orbital radii as or after the disk dissipates; for systems where planets form \emph{in situ}, a full treatment of the first process is necessary.} 

To find the evolution of $J_{2}$ as defined by equation (\ref{eq:scaling}), we use a combination of the numerical models from \citet{Matt2015} for early ($t<1$ Gyr) times, and the \citet{Skumanich1972} $P_{rot} = 2 \pi \Omega^{-1} \propto t^{1/2}$ scaling for late times. The \citet{Matt2015} models use the \citet{Baraffe2015} stellar evolution models combined with stellar wind dynamics to obtain the evolution of $\Omega_{*}$ for stars with a variety of initial conditions (resulting in a range of allowable $\Omega_{*}$ and $J_{2}$ values per time). We use the \citet{Matt2015} models to compute the evolution of $\Omega_{*}$ for the first 1 Gyr of the star's life, then use the Skumanich relation to match boundary conditions at 1 Gyr and model the subsequent evolution of $\Omega_{*}$.
The star transitions from being convectively dominated to radiatively dominated during its first Gyr of life, resulting in a $k_{2}$ constant that decreases from $k_{2} = 0.28$ for a convectively-dominated young star to $k_{2} = 0.014$ for the radiatively-dominated old star. We model the decay using the \citet{Baraffe2015} numerical simulations to find the contribution to the stellar moment of inertia due to the convective and radiative components. We then construct a piece-wise continuous model for $k_{2}$ based on these results, where $k_{2} = 0.28$ for times earlier than $10^7$ years, $k_{2}$ decreases linearly with time between $10^7$ years and $10^8$ years, and $k_{2} = 0.014$ for times later than 1 Gyr. 

\begin{figure}
 \includegraphics[width=3.4in]{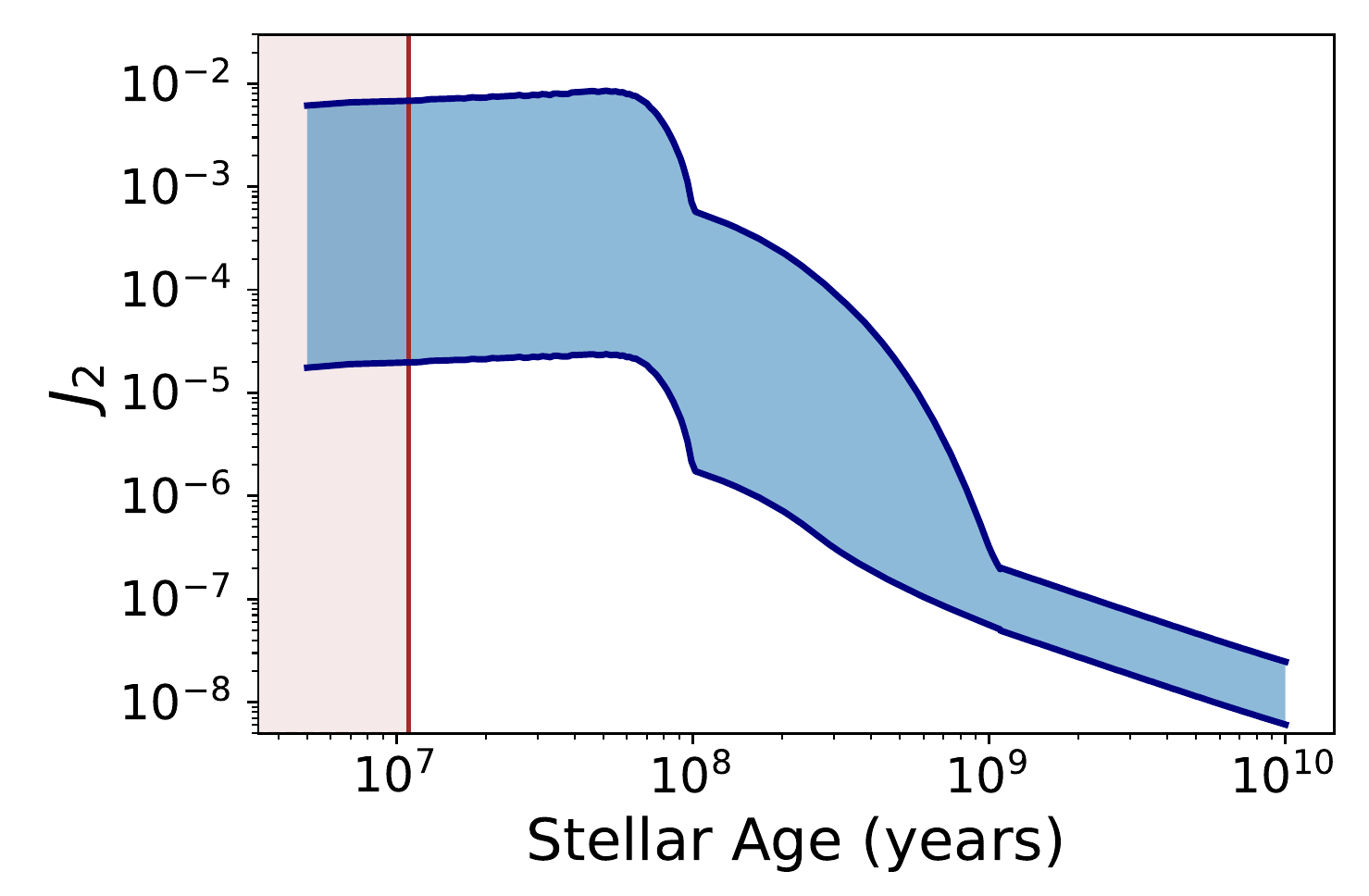}
 \caption{The modeled $J_{2}$ evolution of K2-266 over the stellar lifetime. This scheme uses the stellar models from \citet{Baraffe2015}, $\Omega_{*}$ derived from stellar wind dynamics from \citet{Matt2015} for early times, the scaling relation of \citet{Skumanich1972} for late times, and a time-varying $k_{2}$ defined in the text. The blue range denotes the allowable range of $J_{2}$ at each time, based on the evolution of stars with various initial conditions but with masses within 0.05 $M_{\odot}$ of K2-266's mass (which is 0.686 $M_{\odot}$) from \citet{Matt2015}. }
    \label{fig:j2evolution}
\end{figure}

With $k_2(t)$ specified, we use equation (\ref{eq:scaling}) to compute the $J_{2}$ evolution, finding $\Omega_{*}$ fixing $M_{*}$ and allowing $R_{*}$ to follow the \citet{Baraffe2015} model for a 0.7 $M_{\odot}$ star. The range of allowed $J_{2}$ values are plotted as a function of time in Figure \ref{fig:j2evolution}. We note that because $\dot{J_{2}}$ is adiabatic with respect to all other relevant timescales, neither its actual rate nor exact form matters, as long as the evolution is sufficiently slow. 
For K2-266, we do not have an estimate of the current-day rotation rate. However, for stars of comparable mass and age, the rotation period is expected to be around 20 -- 100 days \citep{Mamajek2008, Angus2015}. 

With the above results in place, we can now compare the snapshot values of $J_{2}$ that we previously found to be sufficient to create the geometry in question to the expected $J_{2}$ evolution for a typical star of this mass. 
We identified in Section \ref{sec:misaligned_axis} values of $J_{2}$ that lead to various classes of behavior. Namely, when $J2 > 10^{-4}$, the USP is coupled to the outer plane and no misalignment is generated; and then when $J_{2} < 10^{-5}$ or so, the observed system geometry can be reproduced.
By an age of 11 Myr, 80 - 95\% of systems of a similar mass to K2-266 are expected to have lost their circumstellar disks \citep{Ribas2015}. We over-plot on Figure \ref{fig:j2evolution} this upper disk lifetime for a star of this approximate mass. For age $\sim10$ Myr when the disk is expected to have dissipated, {the lower envelope of $J_{2}$ values is around $J_2\approx 10^{-5}$, and then subsequently decreases, which would avoid a large mutual inclination within the STIP and create the misalignment between the USP-STIP. We note for stars with higher initial rotation velocities (where $J_2\approx 10^{-2} - 10^{-4}$), the desired geometry might not be produced. If disk damping of inclination could be ruled out, then it might be possible to constrain the early $J_{2}$ of a star based on these arguments}. Note that our uncertainty on the disk lifetime and the intrinsic scatter in the rotation periods of young stars \cite{rebull} prevent a more precise estimate. In any case, {it is likely that if K2-266's initial rotation velocity resided at the lower end of the envelope shown in Figure 5, then K2-266 would have} attained the value of $J_{2}$ needed to produce the required misalignment either right as the disk dissipated or after it was completely gone. This timing of events allows a disk-driven migration mechanism {(defined as any migration mechanism where movement of planets is caused mainly by torques provided by the disk)} to place the planets in their current orbits at an epoch early enough for the stellar $J_{2}$ to be capable of producing a geometry as extreme or more extreme as what we see in the present-day K2-266 system. 

It is important to specify that the reason we consider the $J_{2}$ evolution here is only to ask the following question: at various times in the evolution of the disk, can the $J_{2}$ realistically be what it needs to be to produce the system geometry? The snapshots generated in Section \ref{sec:j2} cannot be connected sequentially to each other in order to model the evolution of inclination over time, as doing so would not preserve the phase space area of the system. An analysis of this problem that does preserve the phase space area and allow the modeling of the evolution of mutual inclinations over billions of years as $J_{2}$ changes will be completed in a companion paper to this work. 

\vspace{10mm}
\section{Effect of an additional (unseen) planetary companion}
\label{sec:companion}

In the previous section, we analyzed the dynamics of the system for given parameters of the validated planets and varying parameters of the host star. However, it is possible --- even likely --- that we have not yet detected all of the planets in this system. Some planets may not be observable in transit, whereas others could have orbital periods long enough that they were not observed during the K2 baseline originally used to discover the known planets \citep{Rodriguez2018}. {The eight epochs of radial velocity observations in \citet{Rodriguez2018} were only enough to barely see K2-266 b, but not to search for additional signals in the system. As a result, the vast majority of parameter space in which an exterior companion could reside in this system is as yet unstudied, leaving ample space for unseen companions.} As a result, in this section we consider the effects that an additional unseen planet would have on the dynamics of the system. 

\subsection{Simulation Setup and Details}

To determine the regime for which an injected companion can reproduce both the co-planarity of the outer transiting planets and the misalignment of the ultra-short-period planet, we perform a Monte Carlo study of the companion parameter space. The inclination of the companion orbit with respect to the starting plane of inner planets is allowed to vary over the range (0,90) in degrees, and the semi-major axis varies over the range (0.1, 2) in AU. The values of $a$ and $i$ are drawn uniformly between these bounds.  We then evolve the system forward in time for $10^{5}$ years under the effect of a $J_{2} = 10^{-7}$ (again, the typical value for a star of this mass and estimated age). All other simulation parameters are as described in Section \ref{sec:sim_params}, and a total of 2500 independent simulations (each with a different, randomly drawn $a,i$ combination) are performed. The orbital inclinations of the known planets are drawn from a Rayleigh distribution with a width of 1.4 degrees.
{In contrast to our earlier examinations of $J_{2}$, here the number of simulations is large enough that we allow these slight typical misalignments between the STIP components. \footnote{Note that in Figure \ref{fig:one_example_companion}, some outer points from the purple and red populations reside in the 'stable' regime, but since we ran enough simulations to robustly differentiate the regions, these outliers do not prevent an understanding of the dynamics.}}
The stellar obliquity is set at $\psi = 0$ degrees for this set of simulations. The eccentricity of our perturbing planet is set to $e  = 0$. For the regime where the perturber has a long period, secular perturbations scale as $(a' \sqrt{1-e^{2}})^{-3}$, so the degeneracy between $a$ and $e$ means that effect of a non-zero eccentricity can be mimicked by changing $a$. The perturber's longitude of ascending node is randomized (since we are not looking for the perturbing planet to transit and the inner system starts axisymmetric, this choice does not affect the results). If the true planet parameters are slightly different from those used, the location of the allowed parameter space for companions will vary, but should be similar in general shape and location. 

\subsection{Parameter space for fixed mass} 

\begin{figure*}
 \includegraphics[width=6.8in]{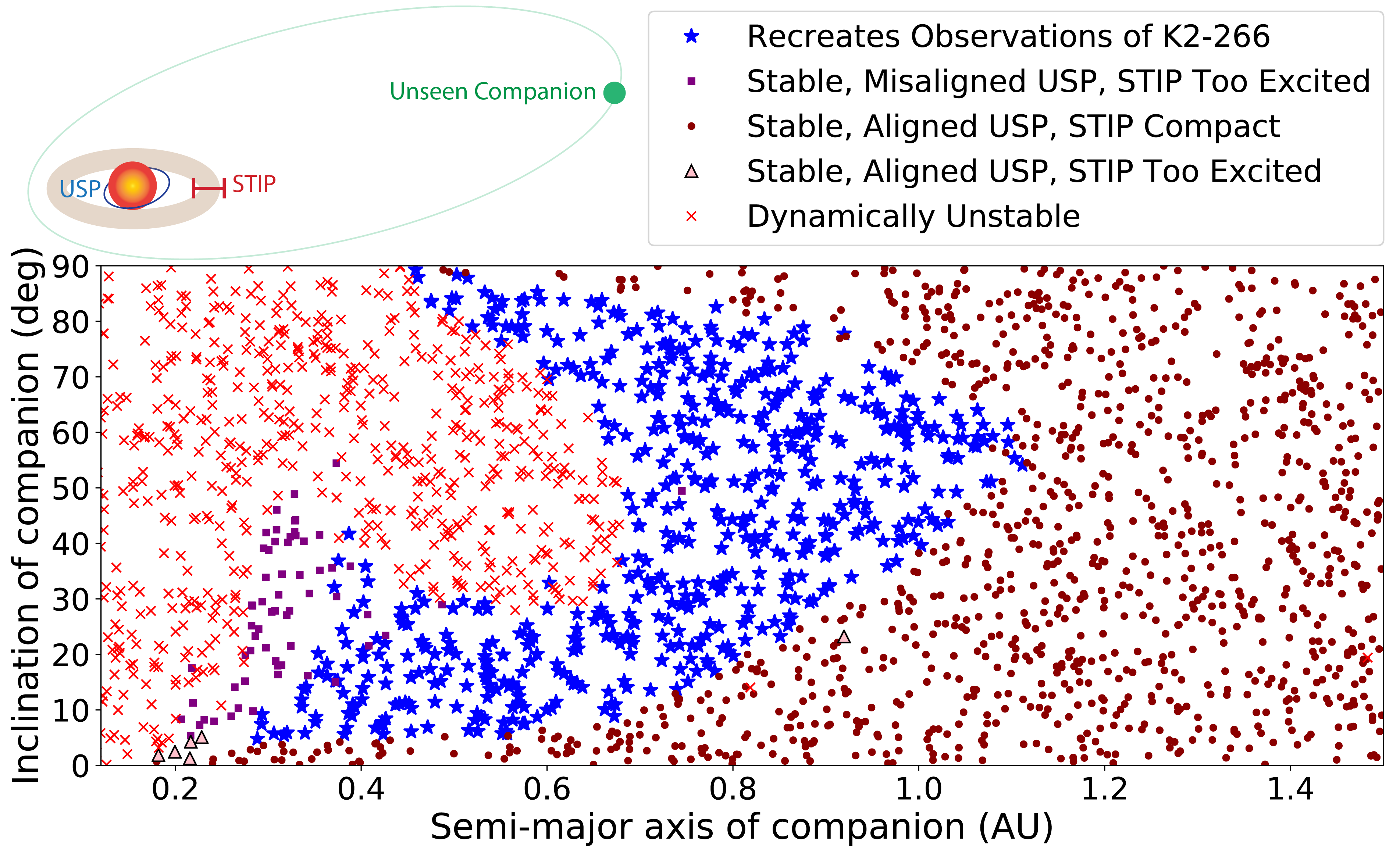}
 \caption{The possible parameter space for a roughly Saturn-mass companion that would cause the observed geometry. Each point in this diagram denotes a different simulation, identical in all ways except for the inclination and semi-major axis of the injected companion. The blue stars denote dynamically stable combinations of orbital parameters that reproduce a median 15 degree misalignment between the USP and the STIP, while points that failed one of the various conditions (dynamically stable as a system, the STIP is well-confined in inclination-space, and the USP is misaligned relative to the STIP) are plotted in colors and point styles described by the plot legend. The acceptable parameter space spans a relatively large range of orbital parameters. }
    \label{fig:one_example_companion}
\end{figure*}

Using the initial conditions outlined above, we have carried out an ensemble of simulations with varying $(a,i)$ for the additional planet, with mass fixed at 100 $M_\oplus$. Figure  \ref{fig:one_example_companion} depicted the resulting possible parameter space that meets the following two specific criteria:
\begin{enumerate}
    \item The outer 3 validated planets (b, c, d) reside in a plane with an inclination dispersion of less than three degrees for more than 30\% of the integration time, 
    \item The median (averaged over time) of the misalignment {in inclination} between the innermost planet, K2-266 b, and the plane of outer transiting planets is greater than 15 degrees. 
\end{enumerate}
The production of these metrics is qualitatively similar to those produced in Figure \ref{fig:part1}, where the simulation results must meet criteria 1 and 2 above in order to count as `reproducing the observations.' {We note that the specific numbers chosen in these criteria are somewhat arbitrary. In particular, the USP-STIP misalignment condition is intended only to produce a system where the observed misalignment is likely, but a 15 degree misalignment is not required for any significant fraction of time, only at the moment the system is observed. These conditions when altered will expand or shrink the region of interest, but the same general shape will hold. }

There are multiple failure modes by which a simulated system may not recreate the observations: the addition of a particular companion may cause the system to go dynamically unstable (which is defined in simulation terms as a close encounter with another object, ejection from the system, or physical collision with the central body), it might cause too great of a misalignment between the STIP components (failing criteria 1), it might fail to recreate the USP-STIP misalignment (failing criteria 2), or it may fail both criteria simultaneously. 

In Figure \ref{fig:one_example_companion}, we show the results of the Monte Carlo parameter space survey, with points color- and marker-coded by which behavior mode they attain (either one of the several varieties of failure or success at reproducing the observations). 
This figure was generated for a companion mass of 100 $M_{\oplus}$ (roughly the mass of Saturn). 

The results of Figure \ref{fig:one_example_companion} show that placing a companion too close to the existing system will lead to dynamical instability, whereas too distant planets will not cause the USP to become misaligned with the STIP. However, there exists an intermediate zone in the $(a,i)$ parameter space (marked by blue stars in the figure) where the companion will drive an initially co-planar system to separate into a decoupled inner USP and an outer STIP. 



\subsection{Dependence of results on companion mass}

In the previous section, we considered a Saturn-mass companion. Although it is not impossible to have Saturn-mass (or larger) planets as part of multi-planet systems around stars with masses $M_\ast$ = 0.6 -- 0.8 $M_{\odot}$ \citep{Johnson2012, Triaud2013}, smaller planets are more common.  We thus need to consider how altering the mass of the companion changes the observed effect. 

\begin{figure}
 \includegraphics[width=3.2in]{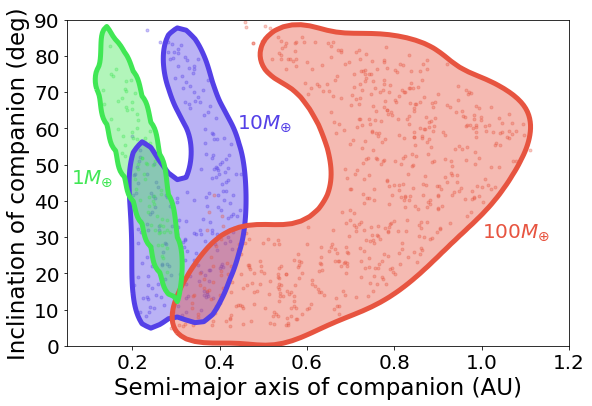}
 \caption{For three masses of perturbing companion (1 $M_{\oplus}$, 10 $M_{\oplus}$, and 100 $M_{\oplus}$), this figure shows the regime of parameter space that allows the inner system to both experience significant misalignments between the USP-STIP planes and keep the STIP well-confined in inclination space. }
    \label{fig:masscompare}
\end{figure}

To answer this question, we ran two additional sets of 2500 simulations each for two additional masses of companion: 1 $M_{\oplus}$ and 10 $M_{\oplus}$. As before, we keep the companion eccentricity at $e=0$ and allow the varying values of semimajor axis $a$ to be the source of initial variation in the potential (as planet mass $m$ is fixed for each set of integrations). These simulations allow us to calculate the portion of ($a,i$) parameter space that recreates the observations for varying masses of the companion. To more easily contrast the allowed parameter space in each system, we compute the two-dimensional Gaussian kernel density estimate (KDE) for each allowed region (excluding all companion parameters that do not recreate the observations and including only those that do) and use Scott's rule as implemented in \texttt{scipy} to automatically compute the bandwidth \citep{Scott2015}. 
We plot the results in Figure \ref{fig:masscompare} overlaid with the raw points from which the KDE is derived (to show that the KDE is a good approximation). 

For all companions plotted here, the extra planet's coupling with K2-266 b (the USP) is weak compared to K2-266 b's coupling with the stellar quadrupole moment, leading to its inclination evolution being incoherent with the outer plane of transiting planets and allowing the creation of the USP-STIP misalignment. 

Figure \ref{fig:masscompare} demonstrates that for our chosen, simplified scenario, a wide range of companion parameters create the observed effect. Even a relatively small planet ($m_P \sim 1 M_{\oplus}$) can cause the secular mode of the outer STIP to decouple from that of the inner USP, if placed in the correct location. 

\vspace{4mm}
\section{Conclusions on K2-266}
\label{sec:conclude}
Some authors have proposed dynamically hot mechanisms for generating these ultra-short-period misaligned planets, such as secular chaos \citep{Petrovich2018}, which requires significant eccentricities on the other nearby planets, {or low-eccentricity migration \citep{Pu2019}, which requires more modest eccentricities.  }
The mechanisms presented in this paper can independently provide the observed misalignment in USP-STIP systems adiabatically, reproducing the observed compact nature of the outer transiting planets, and does not require the excitation of eccentricity for any planets. 

Instead, our proposed mechanisms both function via decoupling the modes of the secular evolution of the outer transiting planets and that of the ultra-short-period planet. 
Both mechanisms can work simultaneously, with the quadrupole potential being more relevant at early times (when the star is less than 100 Myr old for a star of this mass) and the companion being relevant throughout the history of the system. 
These mechanisms could also be at play in other systems hosting ultra-short-period misaligned planets with multiple other transiting planets, but the exact timescales and parameters of the effect in those systems will differ from those derived in this work for K2-266.

\subsection{Implications on general STIP/USP system architectures}
In this work, we have explored the geometry of the K2-266 system. 
We used the specific orbital parameters of this system to initialize our simulations. 
{K2-266's measured parameters suggest that K2-266 must have a fairly low rotation rate (within the envelope of allowed values shown in Figure \ref{fig:j2evolution}, but on the low side) at early times. In this case, the early history of the system may have evolved in the following sequence. First, the host star evolves from $J_2 = 10^{-2}$ to $J_2 = 10^{-5}$ as the planets form in the outer parts of the disk. At around $J_2 = 10^{-5}$, the planets migrate to their final orbital positions as a much faster rate than the $J_2$ decay rate (the planets are migrating at effectively a constant $J_2$). This occurs after the secular resonance would have swept through the system and caused large mutual inclination in the STIP. Then, the star continues to evolve in $J_2$, passing through $J_2 = 10^{-6}$, at which point while the STIP is unaffected, the USP decouples from the STIP and generates a significant USP-STIP misalignment in inclination. As $J_2$ continues to evolve to lower values, the system is locked in that configuration.}

{The exact values given in the previous paragraph are specific to the K2-266 system.}
As shown in previous work \citep{Spalding2016, Li2020}, the dynamics of the interactions between the mutual inclinations of short period planets and the host star obliquity and $J_{2}$ depend  strongly on the orbital periods of the planets in question. As such, to derive specific constraints for other systems (such as TOI-125 or other USP/STIP systems), the analysis in this work should be applied to the specific parameters of those systems.

\subsection{Future work}

Since all stars have changing $J_{2}$ during their early lives, and will often have some degree of stellar obliquity with respect to their planets, the scenario of Section \ref{sec:j2} is expected to arise for many systems. To some degree, such dynamics must be operational in all systems with USP-STIP geometries. In contrast, the scenario developed in Section \ref{sec:companion} is expected to arise less often, as it requires the presence of an additional, unseen planet. Both scenarios have similar observational signatures when viewed at the present epoch (when the stars are old). However, the time evolution of the planetary systems will be different. We can thus test the efficacy of the two scenarios by finding a number of USP-STIP systems of different ages. 

In the case that the current-day geometry is solely due to the evolution of $J_{2}$ and secular dynamics among the known planets only, it might be possible to determine at what time the planet arrived in their current orbital locations. However, in the case that an unseen companion does exist in the system, its presence could complicate this determination. For that reason, a full characterization of the three dimensional architecture of systems of this geometry is important to understanding which mechanism discussed in this work is at play in particular USP-STIP systems. 

It is important to note some caveats in generalizing the above work to other systems. 
First, with respect to creating the USP-STIP misalignment: the analysis in this paper is only complete if we can exclude natal misalignments (generated before the disk dissipates). The appropriateness of this assumption will be assessed and addressed in future work. {In any case, in this work we consider the effect of an additional companion {and} stellar oblateness on a planetary system that started in a roughly co-planar configuration. Future work will address evolution scenarios with different initial conditions. }
Second, with respect to the inclination confinement of the outer STIP: in the present work, we do not consider the effect of the disk damping inclination excitement early in the system. This may allow the STIP to retain a low inclination dispersion even if the effect of $J_{2}$ alone `should' excite the mutual inclinations between the STIP {components}. In this paper, we showed only (Figure \ref{fig:j2evolution}) the approximate $J_{2}$ evolution for a star the size of K2-266. For more massive stars, the decay in $J_{2}$ may occur more quickly, resulting in $J_{2}$ decaying to a level that will not affect the outer planets before the disk dissipates.
The analysis of USP-STIP dynamics by stellar mass is also left for future work.

{Acknowledgements: }
We thank Stephanie Douglas for useful conversations and Sean Matt for sharing what we needed from his models \citep{Matt2015} in a convenient format.
We thank Lynne Hillenbrand for useful conversations. 
J.C.B.~is supported by the Heising-Simons \textit{51 Pegasi b} postdoctoral fellowship.  DF acknowledges support from NASA’s Exoplanet Research Program via grant NASA-NNX17AB93G. AV's work was performed under contract with the California Institute of Technology (Caltech)/Jet Propulsion Laboratory (JPL) funded by NASA through the Sagan Fellowship Program executed by the NASA Exoplanet Science Institute.
We also thank the NASA Exoplanet Science Institute for hosting the 2018 Sagan Exoplanet Summer Workshop, at which this project originated.

{Software: }
This project used the following software: 
pandas \citep{mckinney-proc-scipy-2010}, IPython \citep{PER-GRA:2007}, matplotlib \citep{Hunter:2007}, scipy \citep{scipy}, numpy \citep{oliphant-2006-guide}, Jupyter \citep{Kluyver:2016aa}, Mercury6 \citep{Chambers1999}

\end{document}